\documentclass[]{interact}

\usepackage{epstopdf}
\usepackage[caption=false]{subfig}

\usepackage[numbers,sort&compress]{natbib}
\bibpunct[, ]{[}{]}{,}{n}{,}{,}
\renewcommand\bibfont{\fontsize{10}{12}\selectfont}
\makeatletter
\def\NAT@def@citea{\def\@citea{\NAT@separator}}
\makeatother

\theoremstyle{plain}

\theoremstyle{definition}

\theoremstyle{remark}

\begin{document}

\articletype{ARTICLE TEMPLATE}

\title{Optimization of probe separation distance and cooling time in multi-probe cryoablation technique by arranging probes in triangular and square pattern-A computational approach }

\author{
\name{Gangadhara Boregowda\textsuperscript{a}\thanks{ Gangadhara B. Email: ma19d502@iittp.ac.in} and Panchatcharam Mariappan\textsuperscript{a}}
\affil{\textsuperscript{a}Department of Mathematics and Statistics, Indian Institute of Technology Tirupati, Andhra Pradesh, India}
}
\maketitle

\begin{abstract}
Cryoablation is a minimally invasive and efficient therapy option for liver cancer. Liquid nitrogen was used to kill the unwanted cells via freezing. One of the challenges of cryosurgery is to destroy the complete tumor without damaging the surrounding healthy cells when the tumor is large. To overcome this challenge, multi-cryoprobes were arranged in a polygonal pattern to create a uniform cooling and optimum ablation zone in the tissue. Single, three, and four cryoprobes were placed in the center, triangle, and square patterns to analyze the temperature profile and ablation zone. The results showed that tissue will freeze quickly when cryoprobes are placed in a square pattern. After the treatment of 600 seconds, $99\%$, $96\%$, and $31\%$ of the tumor were killed using four, three, and single cryoprobes, respectively. One of the difficulties in the multi-probe technique is choosing the probe separation distance and cooling time. The volume of the ablation zone, the thermal damage to healthy cells, and the volume of tumor cells killed during the treatment for different probe separation distances of 10 mm, 15 mm, and 20 mm are analyzed. Compared to other settings, a multi-probe technique destroys the entire tumor with the least harm to healthy cells when probes are arranged in a square pattern with a 15 mm space between them.
\end{abstract}


\section{Introduction}
World Health Organization (WHO) estimates that cancer was one of the major causes of death globally in 2019  \citep{bray}. This is because of changes in human lifestyle and the environment. In 2020, there were approximately 19.3 million new instances of cancer and 10 million cancer-related deaths, according to the International Agency for Research on Cancer (IARC) \citep{sung}. Patients with hepatocellular carcinoma (HCC) benefit most from surgical and liver transplantation procedures, which also lower their chance of acquiring new HCC \citep{bruix}. Oncologists advise minimally invasive procedures such as local ablative techniques and liver resection due to patients' habits, the location of the tumor, and a lack of adequate donors. For the treatment of cancer, a variety of ablative methods have been employed, including cryoablation, radiofrequency ablation, and microwave ablation.

For some tumor ablations, cryosurgery is a minimally invasive and efficient therapy option. It is predicated on the idea that unhealthy tissue can be destroyed by subjecting tumor cells to extremely cold temperatures created by cryogenic agents \citep{chua}. The goal is to totally destroy the tumor cells while causing the least amount of cryoinjury to the healthy tissue around them. This sort of treatment, which can be utilized to eliminate skin malignancies, liver, lung, and prostate tumors, is typically carried out concurrently with other cancer-treating techniques like radiotherapy or chemotherapy. Due to its minimum invasiveness, which results in less pain and bleeding, cheaper treatment costs, and shorter hospital stays, cryosurgery can be a suitable substitute for open surgery \citep{deng}. It only has a small cut for the cryoprobe to be inserted.

The cryoprobe is often cooled by convection of liquid nitrogen or argon which circulates inside the probe. Based on its capacity to attain the lowest temperature achievable, the coolant liquid is selected \citep{pasquali}. Liquid nitrogen and argon have boiling points of $-196^\circ$ C and $-186^\circ$ C, respectively. Conduction transfers heat from the cancerous tissue to the cryoprobes during freezing.

Numerous researchers have looked into ways to enhance the cryosurgery process in order to reach the goal of completely destroying tumor cells with the least amount of harm to neighboring healthy cells \citep{malikeh,nazemian, deng,pasquali}. Insufficient and non-uniform freezing of tumor cells may lead to incomplete destruction of the tumor. Many challenges have arisen in this field, such as the functionality of the cryoprobe, effects of vessel size and network on the ablation zone, irregular tumor shape, and creating an ice ball for large tumor \citep{talbot,nazemian}. In the literature \cite{malikeh}, the Author 
 numerically investigated  the effect of vessel size and network on the ablation zone. The literature  \citep{nazemian} demonstrated the shape and formation of the ice ball for large tumors, and it showed that  a single cryoprobe is inefficient in creating the large ice ball for large tumor. An optimization method for planning multi-probe cryosurgery was demonstrated in \citep{giorgi}.

 In this work, multiple cryoprobes are used to create a large ablation zone, and we numerically investigated the shape and size of the ablation zone, and temperature profile in the tissue for different numbers of cryoprobes. The different patterns for cryoprobe locations are proposed for large tumors to optimize the cryoinjury. The tumor nodules have different shapes and sizes in the cancer types, such as breast, lung, and liver. One of the common shapes, which includes the majority of shapes, is the spherical tumor \citep{tehrani,sefidgar, soltani}.  As a result, we assumed the tumor as a sphere with a 25 mm radius. The destruction index $\theta_d$ was used to quantify the cryoinjury or ablation zone \citep{malikeh}. To achieve uniform cooling in the tissue and a spherical ablation zone, the cryoprobes were arranged in a polygonal pattern. 

Selecting the probe separation distance and cooling time, which depends on tumor size, is one of the difficulties of the multi-probe technique. For various probe separation distances and cooling times, we evaluate the volume of the ablation zone, thermal damage to healthy cells, and the volume of tumor cells killed during the treatment. Three different arrangements are used to simulate the results: (1) Cryoprobes are arranged in a triangle and square pattern with a d=10 mm distance between any two probes, (2) Cryoprobes are arranged in a triangle and square pattern with a d=15 mm distance between any two probes, and (3) Cryoprobes are arranged in a triangle and square pattern with a d=20 mm distance between any two probes.
 
 The bio-heat equation with temperature-dependent thermal parameters was used to analyze the temperature profile in the tissue. The bio-heat equation was numerically solved using the Finite Element Method (FEM). The model was discretized and then solved using the open software Gmsh and FEniCS \citep{aln}, respectively. The 3D domain and the bio-heat equation were discretized using the tetrahedral element and Lagrange basis functions, respectively.

\section{Materials and methods }
This section explains the model's geometry, governing equations, boundary conditions, and FEM implementation. The degree of cryoinjury is measured using the destruction index, which ranges from 0 to 1. In order to analyze the temperature profile and ablation zone, the cryoprobes are arranged in a center, triangle, and square configuration. The effect of probe separation distances and cooling times on the ablation zone is analyzed. The problem is discretized and solved using Gmsh and FEniCS, respectively.

\subsection{Geometry of the model}
\begin{figure}[!h]%
\centering
\includegraphics[scale=0.5]{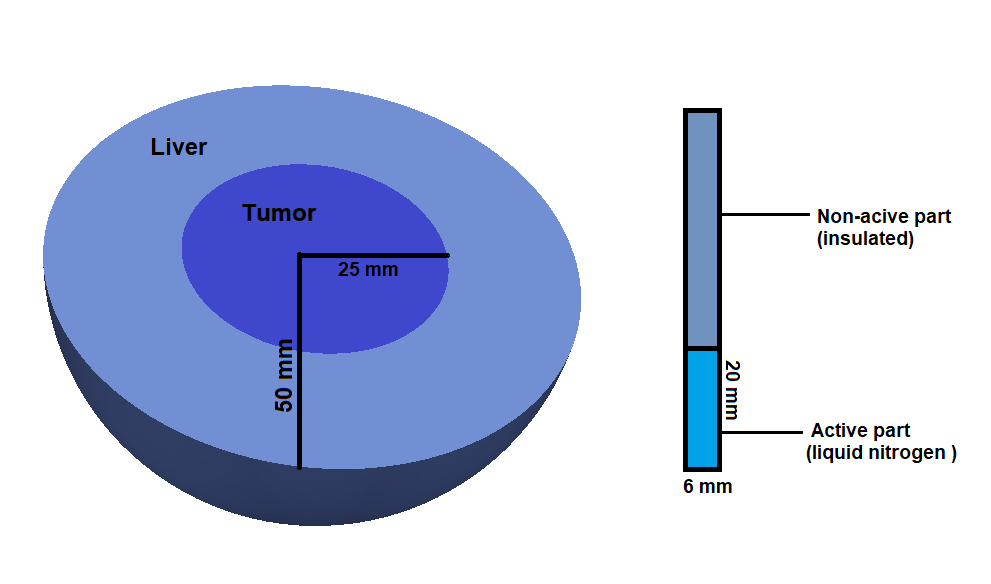 }
\caption{Cross sectional view of the computational domain.} \label{domain}
\end{figure}

The computational domain, which was considered to be a sphere of liver tissue with a radius of 50 mm, is depicted schematically in Figure \ref{domain}. In the middle of the computational domain spherical tumor with a radius of 25 mm was placed. The cryoprobe is represented by a vertical cylinder (6 mm in diameter) with the tip inserted 30 mm into the tumor.  The cryoprobe was divided into two parts. One is an active part, and the other is a non-active part. The length of the active part is 20 mm and it is filled with liquid nitrogen. The non-active part was thermally insulated. The dimension of the probe is based on the literature \citep{chua,malikeh}. The position of cryoprobes in the tissue is explained in subsection \ref{positions}.

\subsection{Governing equation}
In the literature, \citep{nazemian,malikeh}, the domains of the liver, tumor, and blood were taken into consideration to examine the impact of artery size and vascular network complexity on cryosurgery. Since we are investigating the size of the ablation zone for different numbers of antennas, we only take into account the liver and tumor in the computation domain. Moreover, cryoablation has little impact on nearby large blood vessels \citep{niu}. The bio-heat equation was used to analyse the temperature distribution in the liver and tumor \citep{Pennes}.
\begin{equation}\label{bioheat}
    \rho_t c_t \frac{\partial T}{\partial t}=\nabla \cdot( k_t \nabla T)  + \rho_b \omega_{b} c_{b} (T_{b}-T)+Q_{m}  \ \ \text{in} \ \ \Omega 
  \end{equation}
where  $\Omega$ is the computational domain, $\rho_t$ is the tissue density (kg/m$^{3}$), $c_t$  is the specific heat capacity  of the tissue (J/ kg$\cdot ^\circ C$), 
$k_t$ is the tissue thermal conductivity  (W/m$^{\circ} C$), $T$  is the temperature ($^{\circ} C$), $Q_e$ is  the  absorbed  electromagnetic energy (W/m$^{3}$), $ \omega_{b}  $ is the blood   perfusion rate (kg/ m$^{3}\cdot$ s),  $c_{b}$  is the specific heat capacity of blood (J/ kg$\cdot ^\circ C$), $T_{b}$ is the blood  temperature ($^\circ C$), $Q_m$ (W/m$^{3}$) is the metabolic heat rate of the tissue.   

The cells will go through a phase change at the freezing point during freezing, losing the latent heat of freezing in the process. Clinical research has shown that the freezing occurs between $-1\ ^\circ C $ and $-8 \  ^\circ C$. The analysis will be broken down into three different temperature ranges \citep{pasquali,chua}, $-1 \ ^\circ C< T <37 \ ^\circ C$ when cells are unfrozen, $-8 \ ^\circ C < T < -1 \ ^\circ C$ when cells are freezing and $-196 \ ^\circ C<T<-8 \ ^\circ C$ when cells are frozen. The tissue's thermophysical characteristics in the frozen, unfrozen, and freezing stages are as follows \citep{deng}:  
\begin{eqnarray}
k_t&=&\begin{cases}
k_u  & -1 \ ^\circ C <T\\
\frac{k_u+k_f}{2} &-8 \ ^\circ C \leq T \leq -1 \ ^\circ C\\
k_f  &-196 \ ^\circ C < T < -8 \ ^\circ C
\end{cases}\\
C_t&=&\begin{cases}
c_u  & -1 \ ^\circ C <T\\
\frac{c_u+c_f}{2}+\frac{Ql}{\rho_t (T_{u}-T_{l})} &-8 \ ^\circ C \leq T \leq -1 \ ^\circ C\\
c_f  &-196 \ ^\circ C < T < -8 \ ^\circ C
\end{cases}\\
\rho_t&=&\begin{cases}
\rho_u  & -1 \ ^\circ C <T\\
\frac{\rho_u+\rho_f}{2} &-8 \ ^\circ C \leq T \leq -1 \ ^\circ C\\
\rho_f  &-196 \ ^\circ C < T < -8 \ ^\circ C
\end{cases}
\end{eqnarray}
where the subscripts $u$ and $f$ stand for unfrozen and frozen, respectively and $Q_l$ is the latent heat of fusion. The liver and tumor are modeled as homogeneous mediums and the model parameters are listed in Table~\ref{parameters}.

The different perfusion and metabolic rates were considered for the liver and tumor. The metabolic and perfusion rates are higher in the tumor compared to the liver \citep{deng}.

\begin{table}
\tbl{Thermal parameters for the model \citep{chua,deng,mirkhalili}.}
{\begin{tabular}{ccc}
\toprule
Parameter & Units &  Values  \\\\
\midrule
Conductivity of the tissue, $k_t$ & W/m$^{\circ}$C & $k_u=0.5$, $k_f=2$\\
Specific heat of the tissue, $c_t$ &J/ kg$\cdot ^\circ$C &$c_u=3600$, $c_f=1800$  \\
Density of the tissue, $\rho_t$& kg/m$^{3}$ & $\rho_u=1000$, $\rho_f=998$ \\
Conductivity of the blood, $\rho_b$ &W/m$^{\circ}$C & 0.49\\
Specific heat of the blood, $c_b$ &J/ kg$\cdot ^\circ$C & 3600\\
Density of the blood, $\rho_b$&kg/m$^{3}$ & 1050 \\
Blood perfusion rate, $\omega_b$ & ml/s ml & Liver=0.0005, Tumor=0.002\\
Metabolic heat generation, $Q_m$ & W/m$^3$ & Liver=4200, Tumor=42,000\\
Latent heat of fusion, $Q_l$ &MJ/m$^3$ &250\\
Upper limit of the phase transition, $T_u$ & $^\circ$C & -1\\
Lower limit of the phase transition, $T_l$ & $^\circ$C & -8\\
\bottomrule
\end{tabular}}
\label{parameters}
\end{table}

\subsubsection{Boundary conditions}
The following boundary conditions were used to simulate the cryosurgery process:
\begin{itemize}
    \item Thermally insulated boundary condition was applied to the surface of the non-active part of the cryoprobe.
    \begin{eqnarray}
    \Vec{n}\cdot \nabla T=0
    \end{eqnarray}
    \item  The temperature of the surface of the cryoprobe's active part was assumed to be the boiling temperature of the liquid nitrogen ($-196 ^\circ C$).
    \item The temperature at the external boundary of the liver is assumed as body temperature ($37 ^\circ C$).
    \item The initial and blood temperature  are taken as the  body temperature.
\end{itemize}

\subsection{Positions of the cryoprobes in the tissue}\label{positions}
\begin{figure}[h!]%
\centering
\includegraphics[scale=0.8]{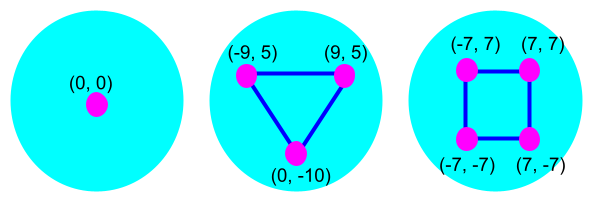}
\caption{cryoprobe locations during the treatment.} 
    \label{position}
\end{figure}

The efficiency of the cryoablation is measured by the ablation zone created during the treatment. For large tumors, a single probe is inefficient in killing all the unhealthy cells in a short time. Therefore, in this research work, we simulated and analyzed the results by using a different number of probes inside the tissue. The three settings are used for the arrangement of cryoprobes in the tissue: (1) a single cryoprobe at the center of the tumor; (2) three cryoprobes located in a triangular pattern with vertices $(-9, 5)$, $(9, 5)$, and $(0, -10)$ in the x-y plane; and (3) four cryoprobes are located in a square pattern with vertices $(-7, 7)$, $(7, 7)$, $(-7, -7)$, and $(7, -7)$ in the $x-y$ plane. To provide a uniform temperature profile in the tissue and a large ablation zone, the probes are positioned in an almost equilateral triangle and square patterns, as shown in Figure \ref{position}.

According to the experiment results \citep{ge}, the tissue's temperature at 10 mm away  from the probe's center after 600 sec was $-50 ^\circ C$. Clinical research has shown that cells will die at $-50 ^\circ C$, regardless of how long they are frozen \citep{gage}.  Therefore, cryoprobes are placed approximately 10 mm away from the center of the tumor domain to produce an optimum ablation zone.

\subsection{Calculation of ablation zone}
The complete death of abnormal cells is necessary for the cryoablation of a tumor. Scientists have demonstrated that necrotic mechanisms begin to function at a temperature of $-5 ^\circ C$, but some undesirable cell survival remains \citep{mazur}. Two methods can be used to guarantee cryo-damage and cell death: (1) The cells will die if the temperature falls below the $-50 ^\circ C$ (cryogenic necrosis temperature $T_{\text{nc}}$) \citep{gage, pasquali}, (2) The cells will die if the temperature stays at $-20 ^\circ C$ (cryogenic damage temperature $T_{\text{nc}}$) for 60 sec ($t_{\text{dc}}$) \citep{cooper}. The first mechanism of destruction takes place close to the cryoprobe, and the second frequently happens in additional tissue that is exposed to lower temperatures.

An index between 0 and 1 was given to various positions to represent the amount of cryo-injury to living tissue. Number one denotes total destruction and indicates that one of the above-mentioned conditions has been met. Zero, on the other hand, denotes an entirely unharmed tissue. Obviously, a higher percentage of tissue damage resulted from the allocated value being closer to 1. The destruction index $\theta_d$ was calculated using the following equations \citep{nazemian,malikeh}:

When the temperature dips below the temperature at which cryogenic damage occurs $T_{\text{dc}}$ for an extended period of time $t_{\text{dc}}$:
\begin{eqnarray}
\alpha_1=\frac{1}{t_{\text{dc}}}\int_{0}^{t}(T<T_{\text{dc}})dt
\end{eqnarray}

When the temperature goes below the cryogenic necrosis temperature $T_{\text{nc}}$:
\begin{eqnarray}
\alpha_2=\int_{0}^{t}(T<T_{\text{nc}})dt
\end{eqnarray}

Combining the two conditions, the overall destruction index is equal to:
\begin{eqnarray}
\begin{cases}
\theta_d=1 & \text{if} \ \ \alpha_2>0\\
\theta_d=\text{min}(1,\alpha_1) & \text{if} \ \ \alpha_2<0
\end{cases}
\end{eqnarray}

\subsection{ FEM implementation to the model }
The bio-heat equation \eqref{bioheat} can be expressed in a variational form by multiplying the test function  $V \in H_0^{1}(\Omega)$ and performing integration by part. The finite element problem is to find $T(x,t) \in  H_0^{1}(\Omega)$  for any $t>0$ such that 
\begin{eqnarray}
    \label{vf1}  	\rho_t c_t \langle \partial_t T,V \rangle +k_t a(T,V)+\rho_b \omega_b c_b \langle T,V \rangle =\langle f, V\rangle, \ \ \forall \ V \  \in H_0^{1}(\Omega) 
\end{eqnarray}
where $\displaystyle \partial_t T=\frac{\partial T}{\partial t},$ $f=\rho_b C_b \omega_b T_b +Q_m \ \in L^2(\Omega)$, $\langle T,V \displaystyle \rangle=\int_{\Omega} T V d\Omega$,
and $$a(T,V)=\int_{\Omega}\nabla T \cdot \nabla V d\Omega$$  

By using the Galerkin approach and basis functions $\{\phi_1, \phi_2,..., \phi_N\}$, the unknown variable is expressed as follows:
\begin{eqnarray} \label{vf2}
T(x,t)=\sum_{i=1}^{N}T_i(t)\phi_i(x)
\end{eqnarray}
where $N$ is the number of nodes and coefficient $T_i(t)$ is functional value of $T(x,t)$ at node $i$.

The  continuous variational form transformed to discrete variational form using equation \eqref{vf2}.

\begin{align}
   \nonumber \rho_t c_t \sum_{i=1}^{N} \partial _t T_i(t)\langle \phi_i, \phi_j\rangle +k_t \sum_{i=1}^{N} T_i(t) a(\phi_i, \phi_j)+\rho_b \omega_b c_b\sum_{i=1}^{N} T_i(t) \langle \phi_i, \phi_j\rangle &=\langle f(t), \phi_j\rangle,
\end{align}
for $j=1,2,3...N$ or in matrix form 
\begin{equation}
\label{vf3} (\rho_t c_t B) \partial_tT(t)+k_t A T(t)+ \rho_b c_b \omega_b B T(t)=F(t)
\end{equation}
where $B=(b_{ij}), \ A=(a_{i j}), \ F=(F_i), \ T=(T_i)$
\begin{align*}
    b_{ij}&=\langle \phi_i, \phi_j \rangle =\int_{\Omega} \phi_i \phi_j d \Omega \\
    a_{ij}&=a(\phi_i, \phi_j)=\int_{\Omega}\nabla \phi_i \cdot \nabla \phi_j d\Omega\\
    F_i&=\langle f, \phi_i\rangle=\int_{\Omega} f \phi_i d \Omega
\end{align*}
Here $B$ and $A$ are mass matrix and stiffness matrix, respectively, which are symmetric and positive definite \citep{johnson} and FEniCS tools \citep{aln} was used to calculate the matrix $B$ and $A$.

By discretizing the time interval  $(0, T)$ into a uniform grid with step size $\Delta t$ and using Euler forward scheme to time derivative term in the  equation \eqref{vf3}, one can find the temperature profile in the tissue at $n^{\text{th}}$ step as follows: 
\begin{align*}
  \rho_t c_t B \left(\frac{T^n-T^{n-1}}{\Delta t}\right)+k_t A T^n +\rho_b \omega_b c_b B T^n&= F^n, \\
  \left[(\rho_t c_t  +\Delta t \rho_b \omega_b c_b) B+ \Delta t k_t A  \right]T^n&=\rho_t c_t B T^{n-1}+\Delta t  F^n.
\end{align*}
MUltifrontal Massively Parallel Sparse direct Solver(MUMPS) \citep{amestoy} was used to solve the above system of equations.

\subsection{Numerical simulation and convergence analysis}
\begin{figure}[h!]%
\centering
\includegraphics[scale=0.5]{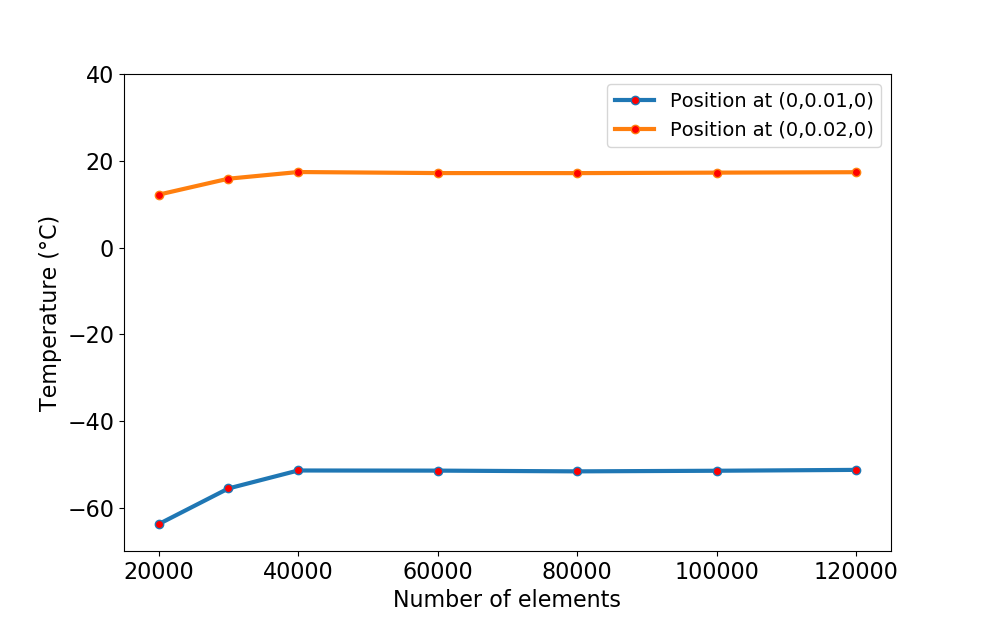}
\caption{Grid convergence} 
\label{meshsensitivity}
\end{figure}

\begin{figure}[h!]%
\centering
\includegraphics[scale=0.4]{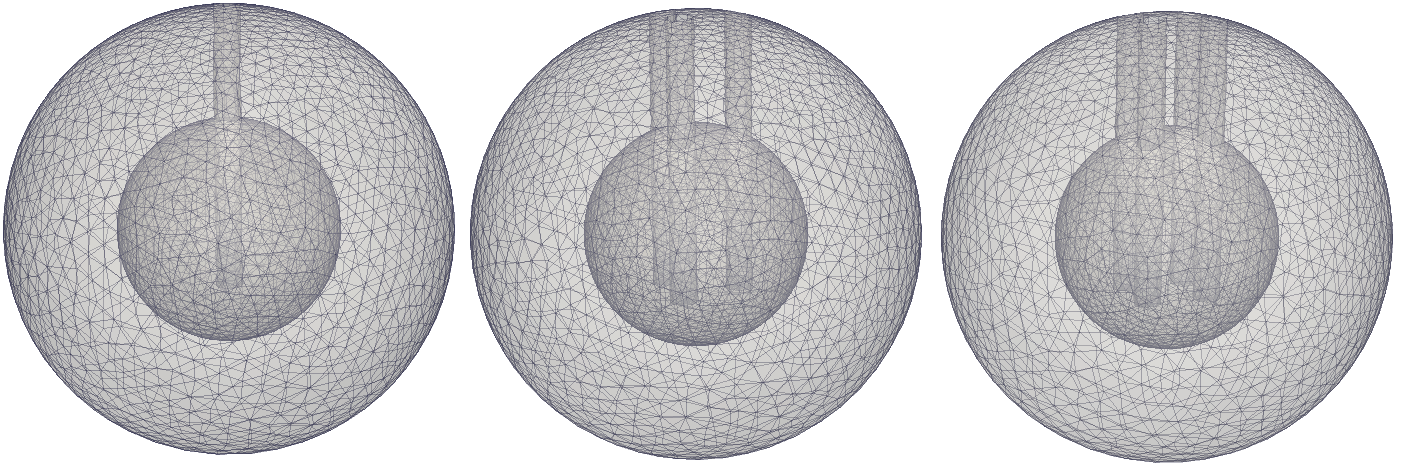}
\caption{Discretization  of Domain with single cryoprobes at the center of the tumor,  three cryoprobes placed in a triangular pattern, and four cryoprobes placed in a square pattern.}
    \label{mesh}
\end{figure}

The bio-heat equation is solved by using FEM via FEniCS. The temporal term in the bio-heat equation was discretized using an unconditionally stable implicit Euler forward scheme. The time step was changed from $0.1$ to $0.4$ with a relative tolerance of $0.0001.$
 The number of elements was altered from $20000$ to $120000$ and the obtained temperature at the points $(0, 0.01, 0)$ and $(0, 0.02, 0)$. Figure \ref{meshsensitivity} shows that after $40000$ tetrahedral elements, the solution is independent of the number of elements. The number of elements for different models is mentioned in Table~\ref{elements}. The domain is discretized using tetrahedral elements via open-source software Gmsh, as shown in Figure \ref{mesh}. A core i5 intel 10$^{\text{th}}$ generation CPU with 8GB memory RAM was used for simulation.

\begin{table}
\tbl{The number of nodes and elements for each model}
{\begin{tabular}{ccc}
\toprule
Model type & Nodes &  Elements  \\
\midrule
Single antenna&8333 & 45489\\
Three antennas in triangular pattern&10837 &59430  \\
Four antennas in square pattern&10671 & 57494\\
\bottomrule
\end{tabular}}
\label{elements}
\end{table}

\section{Results and Discussion}
In this research work, the model of cryosurgery for liver tumors was solved numerically using the FEM with a finite difference scheme for time derivatives.  The effect of the arrangements of the cryoprobes in the tissue on the size of the ablation and temperature profile in the tissue was studied. To verify the simulation method and settings, the numerical results are compared with the experiment results. In this study, we suggested a few patterns for cryoprobe placement in the tissue to create a spherical-shaped ablation zone with homogeneous cooling.

\subsection{Numerical validation}
\begin{figure}[h!]%
\centering
\includegraphics[scale=0.5]{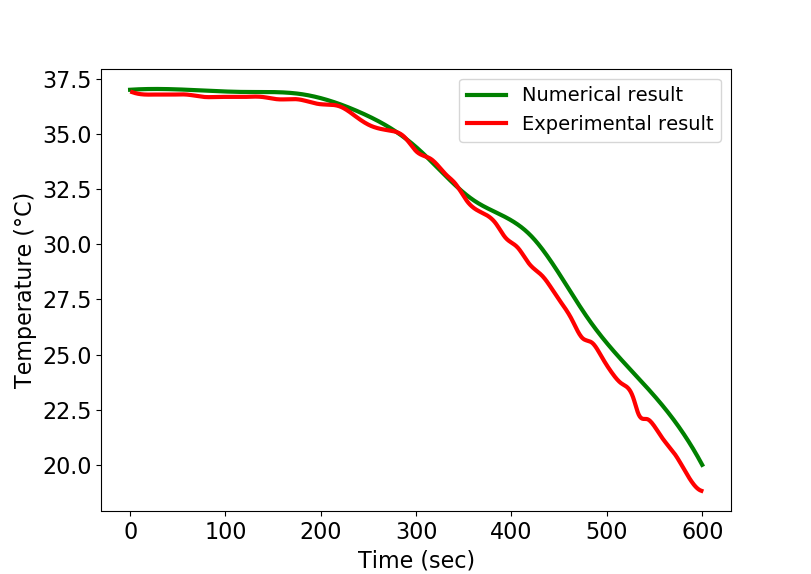}
\caption{Comparison of Numerical results with the experimental result presented in  \citep{ge} at 20 mm away from the center of the probe.} 
    \label{validation}
\end{figure}

Figure~\ref{validation} illustrates the validation of numerical simulation results with experimental results obtained in  \citep{ge}. To compare the numerical results with the experimental results for a 10-minute treatment, the temperature was measured 10 mm away from the cryoprobe's center. The numerical results were in good agreement with experimental results, and the maximum difference was observed at the end of the treatment. The root means square error (RMSE)  between the numerical and experimental results is 0.83 $^\circ$C. We validated the numerical methods, model parameters, and implementation by using RMSE as a reference.  

\subsection{Effect of cryoprobes arrangement on temperature profile and ablation zone}
The cryoprobe is inserted into the tissue and filled with liquid nitrogen at $-196 ^\circ C$ to kill the unwanted cells through freezing. Due to conduction, heat transfers from the liver and tumor towards the cryoprobe during the treatment. As a result, cooling will begin close to the cryoprobe and spread throughout the tissue over time. The temperature of the tumor decreases from $37 ^\circ C$ to $-196 ^\circ C$ as time increases and the necrosis mechanism activates at $-5 ^\circ C$  \citep{mazur}. The goal of this treatment is to totally destroy the unwanted cells with minimum damage to the healthy tissue around the tumor. With less pain, blood, and hospitalization time, cryosurgery is performed by introducing the tiny cryoprobe into the patient's body through a small incision. The treatment will cause structural and phase changes in both healthy and unhealthy cells. As a result, it is considered that density, thermal conductivity, and specific heat capacity are piecewise temperature-dependent functions. In this work, we assumed the computational domain as only the liver and tumor since the cryoablation has less impact on nearby blood vessels \citep{niu}. The computational domain is split into two subdomains because the liver and tumor have different types of material properties \citep{chua}.

\begin{figure}[h!]%
\centering
\includegraphics[scale=0.45]{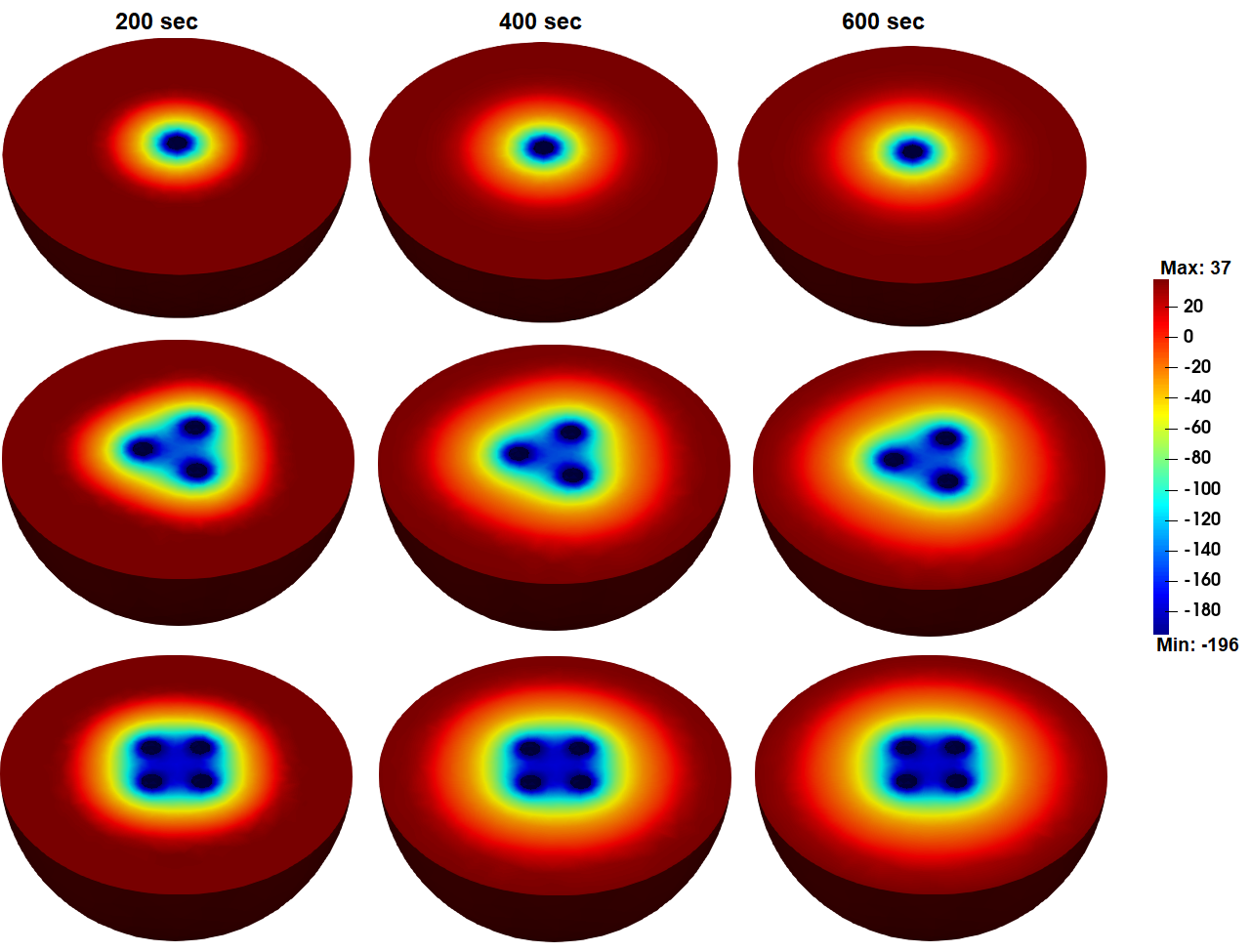}
\caption{The temperature profile ($^\circ$C) in the tissue using cryoprobes in a different pattern.} 
    \label{temp_center}
\end{figure}

One of the challenges in cryoablation is forming a large ice ball for a large tumor without damaging healthy cells. The single cryoprobe is inefficient for creating large ice balls to destroy the large tumor. At the same time, introducing a more number of cryoprobes into the tissue without proper arrangement may kill the healthy cells surrounded by the tumor. Since the spherical shape covers most of the tumor shape \citep{tehrani}, the cryoprobes are placed in the tissue in a polygonal pattern by keeping approximately the same length between any two cryoprobes to create the ablation zone in a spherical shape. The results are simulated for three different configurations: (1) a single cryoprobe positioned at the tumor's center; (2) three cryoprobes arranged in a triangular pattern, each about 10 mm from the tumor's center; and (3) four cryoprobes arranged in a square pattern, each about 10 mm from the tumor's center. The positions of the cryoprobes in a triangular pattern are $(-9, 5)$, $(9, 5)$ and $(0, -10)$ in the x-y plane. Similarly, in a square pattern, positions for cryoprobes are $(-7, 7)$, $(7, 7)$, $(-7, -7)$, and $(7, -7)$. The length of the square and triangular sides can be adjusted according to the size of the tumor.

\begin{figure}[h!]%
\centering
\includegraphics[scale=0.45]{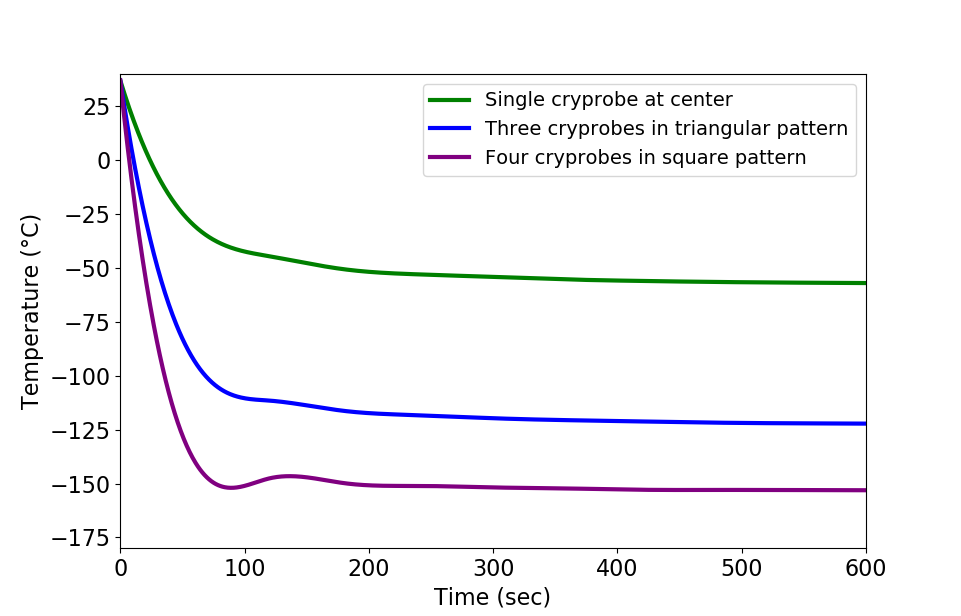}
\caption{The temperature distribution at the position $(0, 0.01, 0)$ during the treatment.  }
    \label{temp_at_10}
\end{figure}

\begin{figure}[h!]%
\centering
\includegraphics[scale=0.45]{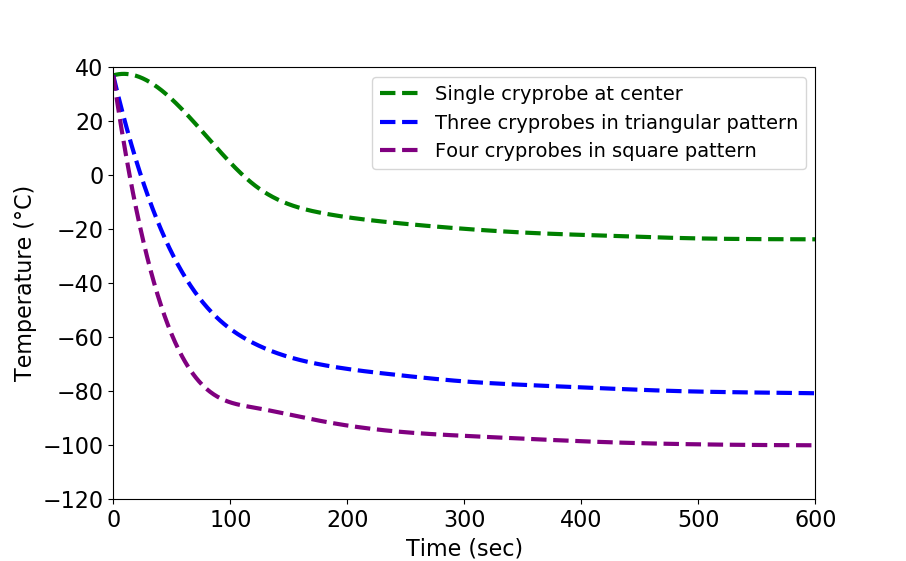}
 \caption{The temperature distribution at the position $(0, 0.015, 0)$ during the treatment.  }
    \label{temp_at_15}
\end{figure}

Figure ~\ref{temp_center} illustrates the temperature distribution in the tissue at $200$ sec, $400$ sec, and $600$ sec using single, three, and four cryoprobes in the center, triangle pattern, and square pattern, respectively. As time passes due to conduction, the temperature in the tumor begins to drop from $37 ^\circ C$  to a lower temperature. The tissue temperature near the cryoprobe is $-196 ^\circ C$, and it increases to a body temperature of $37 ^\circ C$ as it moves away from the cryoprobe. The formation of ice balls in the tissue increases as time increases. The size of the ice ball is a little more when cryoprobes are placed in a square pattern rather than in a triangle pattern. The size of the ice ball is very small when a single cryoprobe is used in cryosurgery. Therefore, using a single cryoprobe for a large tumor may not kill all the unhealthy cells in the tissue. The tissue temperature decreases fast in the first $100$ sec, then it remains constant for the rest of the time, as shown in Figure \ref{temp_at_10} and Figure \ref{temp_at_15}. The temperature was monitored at two positions $(0, 0.01, 0)$ and $(0, 0.015, 0)$ in the tissue during the treatment of 600 sec. When one, three, or four cryoprobes are employed in the procedure, the tissue temperature at $(0, 0.01, 0)$ is reported as $-57 ^\circ C, -122.2 ^\circ C$, and $-153.1^\circ C$, respectively. Similar to this, the temperature is reported as $-23.7 ^\circ C
, -80.72 ^\circ C$, and $-100 ^\circ C$  at $(0, 0.015, 0)$. The tissue will freeze quickly when cryoprobes are placed in a square pattern during the treatment compared to a triangular pattern. Using a single cryoprobe in treatment will take more time to freeze the tissue compared to the other two cases.

\begin{figure}[h!]%
\centering
\includegraphics[scale=0.45]{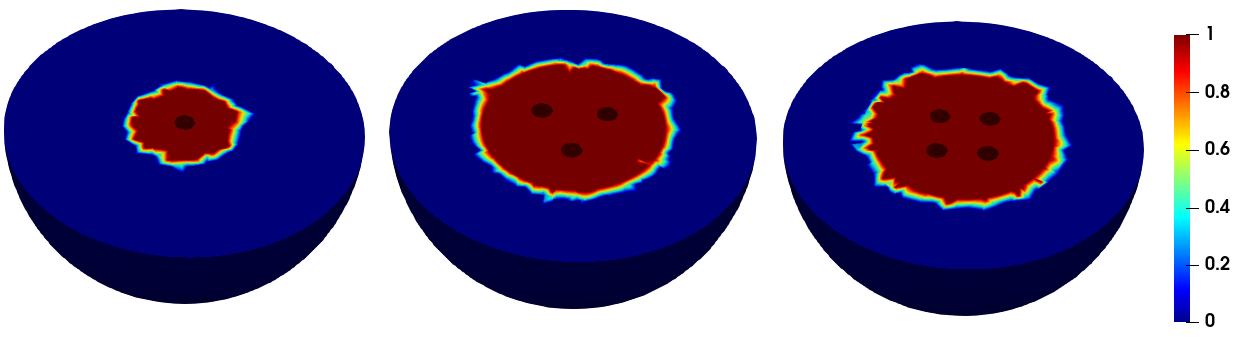}
  \caption{ Destruction index for the treatment of  600 sec.}
    \label{destruction_index}
\end{figure}

\begin{figure}[h!]%
\centering
\includegraphics[scale=0.4]{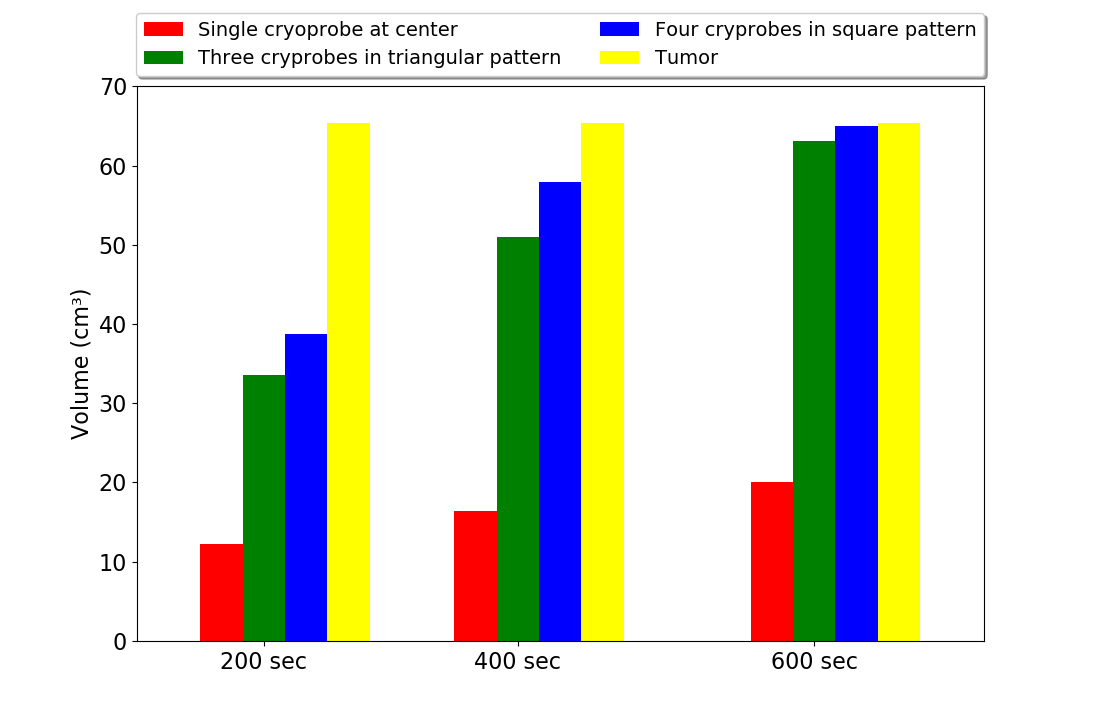}
  \caption{Comparison of ablation volumes obtained in all three settings with tumor volume.}
    \label{volume_comparision}  
\end{figure}

To guarantee cell death in the tissue, two strategies were used \citep{pasquali,cooper}.
One involves cooling the cell to $-50 ^\circ C$, while the other involves keeping it at $-20 ^\circ C$ degrees for $60$ seconds. Based on the above two techniques, the destruction index was defined between $0$ and $1$ to find the ablation zone \citep{nazemian,malikeh}. If the destruction index is close to $1$, then the percentage of cell injury is higher. For dead and undamaged cells, a destruction index is defined as $1$ and $0$, respectively. Figure~\ref{destruction_index} illustrates the destruction index for different arrangements of the cryoprobes. The destruction index is $1$ near the cryoprobes and it decreases to $0$ when it moves away from the cryoprobes, which means cells are dead only around the cryoprobes.
\begin{figure}[h!]%
\centering
\includegraphics[scale=0.45]{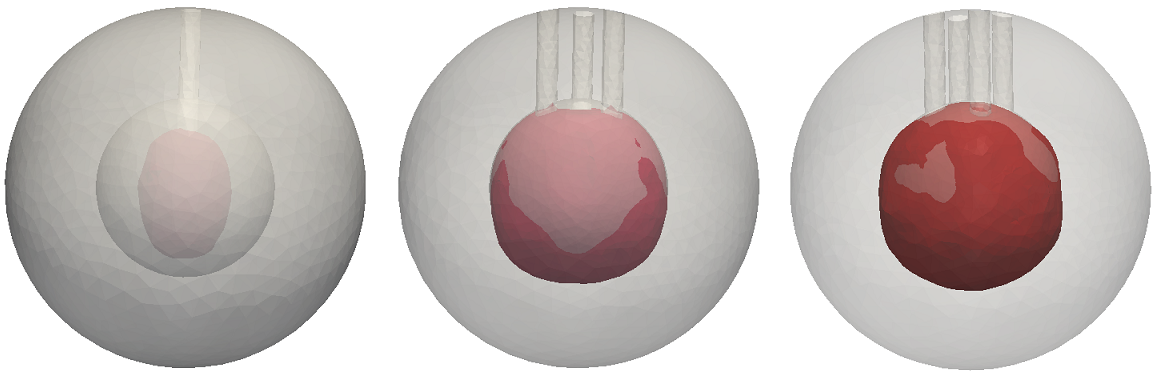}
   \caption{The ablation zone (red color) using a single cryoprobe at the center of the tumor, three cryoprobes located in a triangle pattern, and four cryoprobes located in a square pattern for the treatment of $600$ sec.}
    \label{ablatin}
\end{figure}
One of the challenges in ablation treatments is to kill the complete tumor cells in a short time. When the diameter of the volume is more than 4 cm, using a single cryoprobe in the treatment fails to kill all the tumor cells in a short time. The shape of the ablation zone is approximately spherical when the cryoprobes are located in the square and triangular patterns, and the ablation zone is prolate in shape when a single cryoprobe is placed at the center, as shown in Fig.~\ref{ablatin}. After the treatment of 600 sec, $99\%$, $96\%$, and $31\%$ of the tumor were killed using four, three, and single cryoprobes, respectively. The volume of the ablation zone increases as time increases, as shown in Figure \ref{volume_comparision}. The maximum number of tumor cells is killed when four cryoprobes are placed in a square pattern.   

\subsection{Effect of separation distance and cooling time on the ablation zone for multi-probe cryoablation}
One of the difficulties in the multi-probe technique is choosing the probe separation distance and cooling time, which depend on tumor size. In this work, we analyze the volume of the ablation zone, thermal damage to healthy cells, and the volume of tumor cells killed during the treatment for different probe separation distances and cooling times. Three different arrangements are used to simulate the results: (1) Cryoprobes are arranged in a triangle, and square pattern with a d=10 mm distance between any two probes, (2) Cryoprobes are arranged in a triangle and square pattern with a d=15 mm distance between any two probes, and (3) Cryoprobes are arranged in a triangle and square pattern with a d=20 mm distance between any two probes. 

\begin{figure*}[h!]
\subfloat[]{
    	\begin{minipage}[c][1\width]{0.33\textwidth}
    	\centering
    		\includegraphics[width=5cm,height=5.5cm]{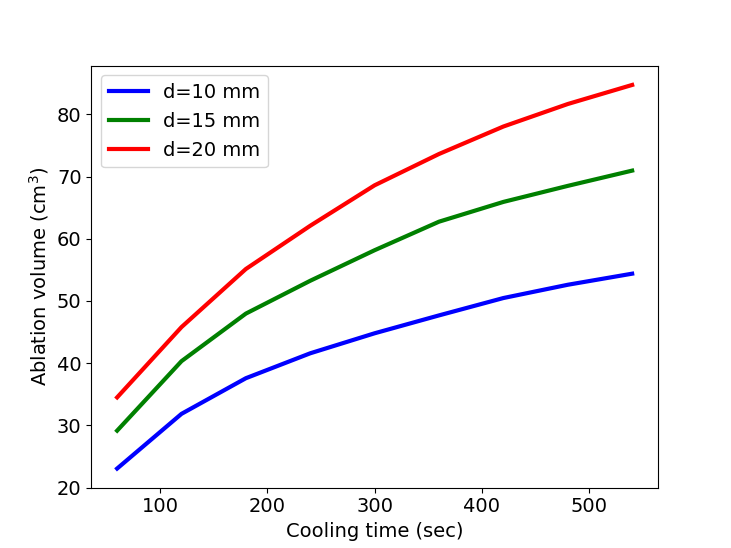}
    		\vspace{0.3cm}
    	\end{minipage}}
    	\subfloat[]{
    	\begin{minipage}[c][1\width]{0.33\textwidth}
    		\centering
    	\includegraphics[width=5cm,height=5.5cm]{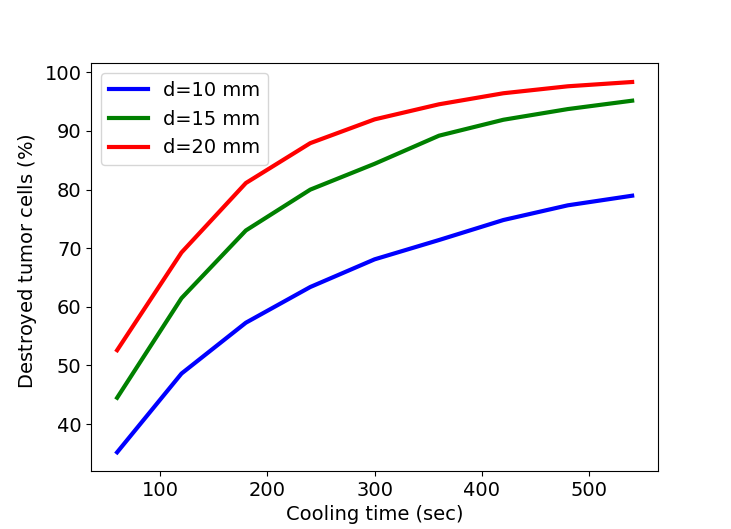}
    		\vspace{0.3cm}
    	\end{minipage}}
    	\subfloat[]{
    	\vspace{-0.5cm}
    	\begin{minipage}[c][1\width]{0.33\textwidth}
    		\centering
    	\includegraphics[width=5cm,height=5.5cm]{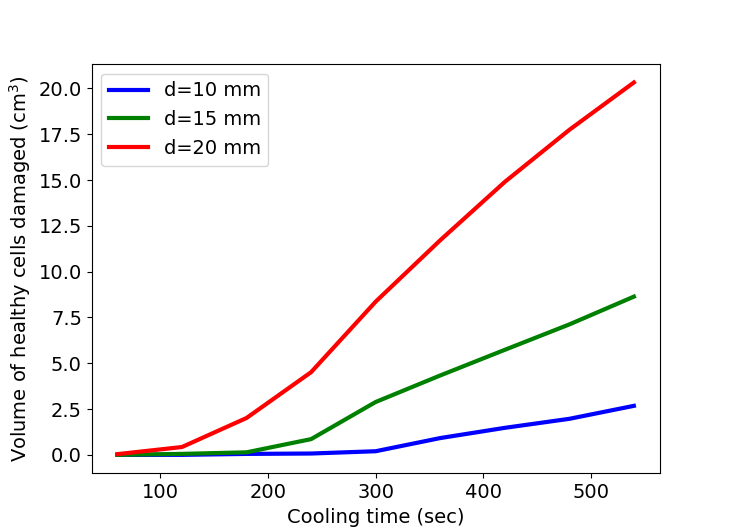}
    		\vspace{0.3cm}
    	\end{minipage}}
     \caption{Cells responses: when three probes  are located in a triangle pattern for different probe separation distances (d) during the treatment }
     \label{tri_sep}
    \end{figure*}

\begin{figure*}[h!]
\subfloat[]{
    	\begin{minipage}[c][1\width]{0.33\textwidth}
    	\centering
    		\includegraphics[width=5cm,height=5.5cm]{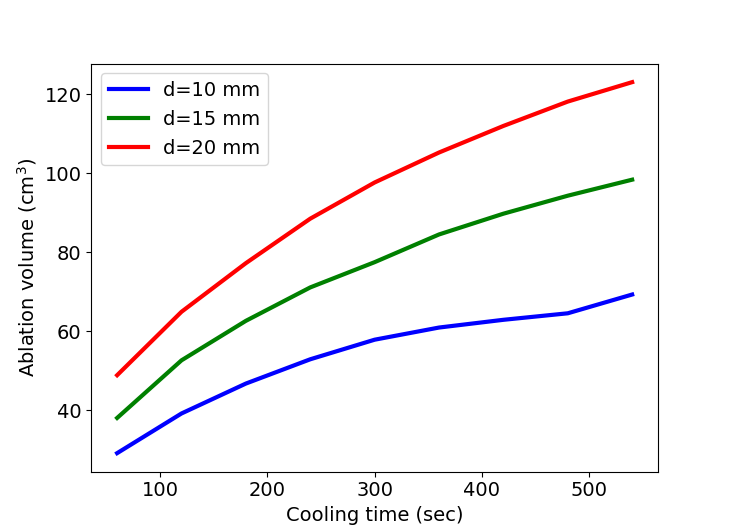}
    		\vspace{0.3cm}
    	\end{minipage}}
    	\subfloat[]{
    	\begin{minipage}[c][1\width]{0.33\textwidth}
    		\centering
    	\includegraphics[width=5cm,height=5.5cm]{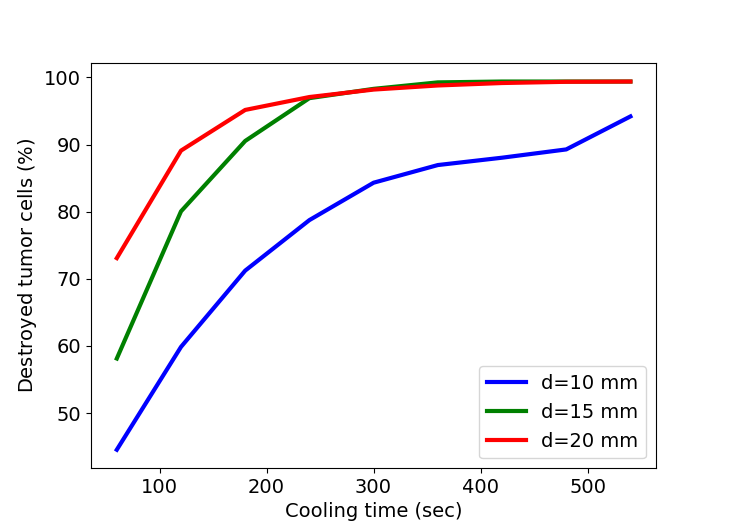}
    		\vspace{0.3cm}
    	\end{minipage}}
    	\subfloat[]{
    	\vspace{-0.5cm}
    	\begin{minipage}[c][1\width]{0.33\textwidth}
    		\centering
    	\includegraphics[width=5cm,height=5.5cm]{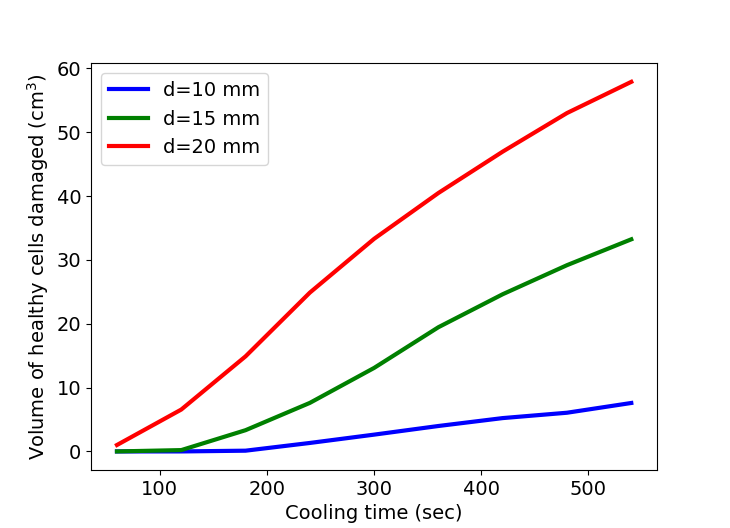}
    		\vspace{0.3cm}
    	\end{minipage}}
      \caption{Cells responses: when four probes  are located in a square pattern for different probe separation distances (d) during the treatment }
     \label{squ_sep}
    \end{figure*}

The ablation volume and thermal damage increase with the separation distance and cooling time.  The multi-probe technique kills $79 \ \%$, $95 \ \%$, and $98\ \%$ of the tumor for a 10 minutes cooling period when cryoprobes are spaced 10 mm, 15 mm, and 20 mm from one another in a triangle arrangement. Similarly, when cryoprobes are separated by 10 mm, 15 mm, and 20 mm in a square pattern, the multi-probe technique kills $94 \ \%$, $100 \%$, and $100 \%$ of the tumor for 10 minutes cooling time, respectively. Multi-probe techniques fail to kill the complete tumor when cryoprobes are separated by 10 mm and 15 mm in a triangular pattern. It kills almost the entire tumor with thermal damage of 20 cm$^3$ when cryoprobes are separated by 20 mm, as shown in Figure~\ref{tri_sep}. Multi-probe cryoablation technique kills almost tumor when probes are separated by 15 mm and 20 mm from each other in a square pattern with thermal damage of 19 cm$^3$ and 40 cm$^3$, respectively, for 360 seconds, as shown in Figure~\ref{squ_sep}. When probes are arranged in a square pattern with a 15 mm spacing between them, a multi-probe approach optimizes thermal damage and cooling time. Probe separation distance and cooling time depend on the tumor shape and size. The relation between the separation distance, cooling time, tumor size, and shape for the multi-probe ablation technique is left for future investigation.

\section{Conclusion}
The liver and tumor were modeled in three dimensions for computational analysis. The thermo-physical properties were applied to liver and tumor tissues in three different phases: unfrozen, freezing, and frozen. Single, three, and four cryoprobes were arranged in the center, triangle, and square patterns, respectively, in order to study the impact of cryoprobe arrangement on the temperature profile and ablation zone. This work shows that placing cryoprobes in a polygonal pattern will freeze the tissue uniformly and create a large ablation zone in a spherical shape. Using four, three, and one cryoprobe, respectively, $99\%, 96\%$, and $31\%$  of the tumor were destroyed after the $600$ second treatment. And also, this study analyzes the volume of the ablation zone, thermal damage to healthy cells, and the volume of tumor cells killed during the treatment for different probe separation distances and cooling times. When probes are arranged in a square pattern with a 15 mm spacing between them, a multi-probe technique kills the complete tumor with minimum damage to healthy cells in a short cooling time compared to other settings. Therefore, this study helps physicians to arrange cryoprobes during the treatment of a large tumor.

\section*{Acknowledgement}
None.
\section*{Disclosure statement}
No potential conflict of interest was reported by author(s).
\section*{Funding}
None.

\end{document}